\documentclass[apl,preprint,showpacs]{revtex4}
\usepackage{graphicx}
\newcommand{\laprs}{La$_{0.175}$Pr$_{0.45}$Ca$_{0.375}$MnO$_{3}$ }
\begin{document}
\title{Complementarity of perturbations driving insulator-to-metal transition in a charge ordered manganite.}

\author{S. Chaudhuri and R. C. Budhani}

\affiliation{Condensed Matter Low Dimensional Systems Laboratory,
Department of Physics, Indian Institute of Technology Kanpur,
Kanpur - 208016, India}


\begin{abstract}Modulation of charge carrier dynamics and hence electrical
conductivity of solids by photoexcitation has been a rich field of
research with numerous applications. Similarly, electric and
magnetic field assisted enhancement of conductivity  are of
fundamental importance and technological use.  Hole doped
manganites of the type (A$_{1-x}$B$_{x})$MnO$_{3}$, where A and B
are rare and alkaline earth metals respectively have the
distinction of showing all three effects. Here we establish the
complementarity of the electric, magnetic and photon fields in
driving an insulator-metal transition in epitaxial thin films of
\laprs whose electrical ground state is insulating. Both pulsed
and CW lasers cause a giant photon flux dependent enhancement of
conductivity. It is further observed that electric and magnetic
fields  trigger the persistent enhancement of conductivity whose
magnitude can be accentuated by application of these fields in
parallel.
\end{abstract}

\maketitle

\section{Introduction}
There is   growing evidence indicating that the external
perturbation induced colossal response of the electrical and
magnetic states in doped manganites is due to the collapse of a
phase, coexisting and competiting with other phases, into an
energetically more favourable state
\cite{salamon,tokura,coey,rao,mathur}. The  external stimuli used
to drive the phase collapse are mostly electric \cite{asamitsu}
and magnetic fields \cite{tomioka}, pressure \cite{hwang} and
internal parameters like doping \cite{urushibara} and disorder
\cite{akahoshi}. Other external stimuli that appears promising to
control the phases in manganites is irradiation with optical
photons \cite{tokura2,miyano,gilabert,cauro2,takubo}, x-rays
\cite{kiryukhin} and electrons \cite{hervieu}. A photo-induced
insulator-metal transition (PIMT) was observed first by Miyano
\textit{et al.} \cite{miyano} in a Pr$_{1-x}$Ca$_{x}$MnO$_{3}$
crystal when the sample was biased with strong electric field in
its charge ordered (CO) state. The induced metallic character was
persistent as long as the bias voltage and sample temperature were
held constant. These observations were attributed to light induced
melting of the CO phase which presumably results in  opening of
metallic percolating channels in the sample. However, these
experiments were performed on thick crystals and issues such as
non-uniformity of photon flux and possible heating of the sample
at high light intensity prohibit a quantitative understanding of
the PIMT. Some of these difficulties can be eliminated if thin
epitaxial films of thickness equal to penetration depth of light
are investigated. Furthermore, for any possible technological
application of PIMT, thin films are desired. Takubo \textit{et
al.} \cite{takubo} first demonstrated a persistent change in
resistivity of Pr$_{0.55}$(Ca$_{1-y}$Sr$_{y}$)$_{0.45}$MnO$_{3}$
epitaxial thin films upon pulsed ($\approx$ 7ns) laser irradiation
without an assisting electric field. However unlike the case of
Pr$_{1-x}$Ca$_{x}$MnO$_{3}$ crystals, this transition occurred on
irradiating the sample with a pulsed laser at a temperature (77 K)
which is only a few degrees higher than the temperature
\textit{T$_{IM}$} ($\approx$ 75 K), where an insulator-to-metal
(I-M) transition occurs on cooling the sample even in dark. While
these materials can be driven to a metallic state with other
perturbations as well, such as electric and magnetic fields
\cite{tokura,budhani,rairigh}, an equivalence of such
perturbations needs to be established.

In this letter we show a giant and  persistent photo-induced
effect in \laprs thin films upon illumination with both CW and
pulsed lasers of wavelengths spanning the range 325 to 632 nm. Our
results display several unique features as compared to earlier
works on photoeffects in charge-ordered manganites. Firstly,
unlike the Pr$_{0.55}$(Ca$_{1-y}$Sr$_{y}$)$_{0.45}$MnO$_{3}$
sample
 \cite{takubo} these films do not undergo an insulator-to-metal
transition on cooling down to $\approx$ 10 K. However a single 6 ns
pulse of 532 nm radiation triggers a $\approx$ 6 orders drop in
resistance over a broad range of temperature. Secondly, we note that
even a CW laser radiation or a simple white light can trigger the
transition albeit with some time delay. Third, contrary to the
earlier report \cite{takubo} the photon induced low resistance state
is irreversible under CW illumination and the virgin state of the
sample is recovered only after cycling it through \textit{T$_{CO}$},
the charge ordering temperature. Lastly, we note that a similar
change in the conductivity can be triggered by a strong voltage bias.
The fact that the threshold voltage for switching can be lowered by a
magnetic field establishes a unique complementarity of perturbations
which trigger the I-M transition in CO-manganites.

\section{Experimental}
Thin films of thickness $\approx$ 200 to 300 nm of \laprs (LPCMO)
were grown at a deposition rate of $\approx$ 1.6 \AA/s on (110)
oriented single crystals of NdGaO$_{3}$ (NGO)  using pulsed laser
deposition (PLD) technique. A KrF excimer laser operated at 10 Hz
with an areal energy density of 2 Jcm$^{-2}$/pulse on the surface
of a stoichiometric sintered target of LPCMO was used for
ablation. These films were then patterned into a 1 mm wide strip
using Ar$^{+}$ ion milling followed by deposition of silver pads
separated by 50 $\mu$m gap for making electrical contacts. The
small bridge dimension (50 $\times$ 1,000 $\mu$m$^{2}$) ensured
uniform illumination of the sample by the laser beam. The
resistance was measured in a standard four probe configuration by
biasing the sample either in a constant voltage (CV) or a constant
current (CC) mode. For the CC measurement a current source [Model
- Kiethley 220] was used in conjunction with a nanovoltmeter (
Input Impedance $\approx$ 1G$\Omega$, Model - Kiethley 2182 ). For
CV measurements, the sample in series with a 100 $\Omega$ standard
metal film resistor, was biased using a  DC power supply [Model -
HP6634B] and its resistance was calculated by measuring the
voltage drop across it and the metal film resistor using a dual
channel  nanovoltmeter (Input Impedance $\approx$ 1 G$\Omega$,
Model - HP34420A). If not mentioned explicitly, the bias current
(CC mode) or the bias voltage (CV mode) are  1 $\mu$A or 1 V
respectively.  The current and the voltage were not applied
continuously but in the form of short pulses of  few hundred
milliseconds duration. The time interval between the pulses was
$\approx$ 2 secs. For photo-illumination we have used the 325 nm
(3.81 eV) and 442 nm (2.8 eV) radiation from a He-Cd CW laser
[Kimmon IK 5552R-F], 632.8 nm (1.96 eV) CW radiation from a He-Ne
laser and 532 nm (2.33 eV) laser pulses from a Q-switched Nd:YAG
laser [ Quantel Model - ULTRA].

\section{Results and Discussions}

\begin{figure}
\includegraphics [width=12cm]{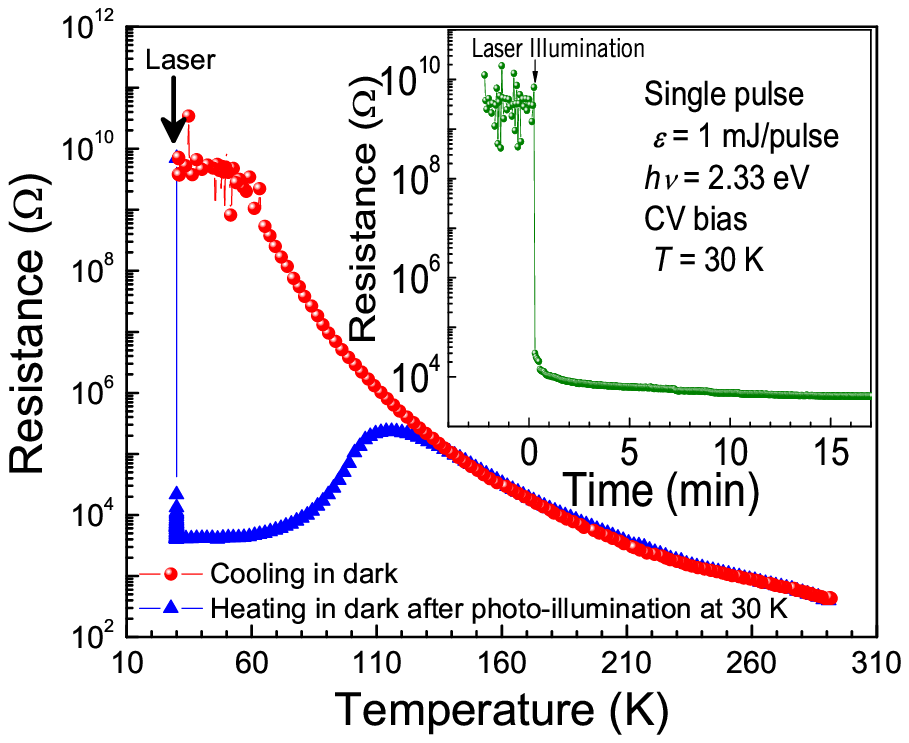}
\caption{(Colour online)Temperature dependence of resistance
measured in the Constant Voltage (CV) mode during cooling in dark
(filled circles). At 30 K the sample was irradiated with 30 pulses
of 532 nm (\textit{h$\nu$} = 2.33 eV) radiation from an Nd:YAG
laser. The irradiation  triggers $\approx$ 6 orders of magnitude
drop in resistance as indicated by the sharp drop marked by the
arrow. After photo-illumination the sample was heated back to room
temperature in dark (filled triangles). The resistance after
illumination indicates a metallic regime between 30 K to 120 K.
Inset of the figure shows the time dependence of the sample
resistance after firing a single shot of 532 nm pulse of energy 1
mJ indicating that the resistance switching is a single shot
event} \label{fig1}
\end{figure}

In Fig.~\ref{fig1} we show the temperature (\textit{T}) dependence
of the sample resistance (\textit{R}), measured in the CV mode,
during cooling (filled circles). The resistance is thermally
activated with a temperature dependent activation energy down to
60 K. For \textit{T} $<$ 60 K, the sample resistance exceeds the
input impedance of the nanovoltmeter. At 30 K, the sample was
irradiated with 30 pulses ($\approx$ 6 ns) of 532 nm (photon
energy \textit{h$\nu$} = 2.33 eV) radiation at 5 Hz each of energy
($\epsilon$) 1 mJ/pulse. The sample resistance dropped by
$\approx$ 5 orders of magnitude upon the incidence of the first
pulse. The subsequent pulses reduce the resistance only marginally
from 3$\times$10$^{4}$ $\Omega$ to 4$\times$10$^{3}$ $\Omega$. The
electrical state of the material after photo-irradiation  is
metallic as can be seen from the positive d\textit{R}/d\textit{T}
of the resistance curve on warming the sample (filled triangles).
Moreover, the photoinduced metallic state persist even after
switching off the electric field. In the inset of Fig.~\ref{fig1}
we show the result of a similar experiment where the sample was
cooled to 30 K from room temperature (300 K), irradiated with a
single laser pulse and the resistance was monitored as a function
of time under isothermal conditions. The resistance drops almost
instantaneously upon the incidence of the pulse followed by a much
slower diffusive type of enhancement in the conductivity.

\begin{figure}
\includegraphics [width=8cm]{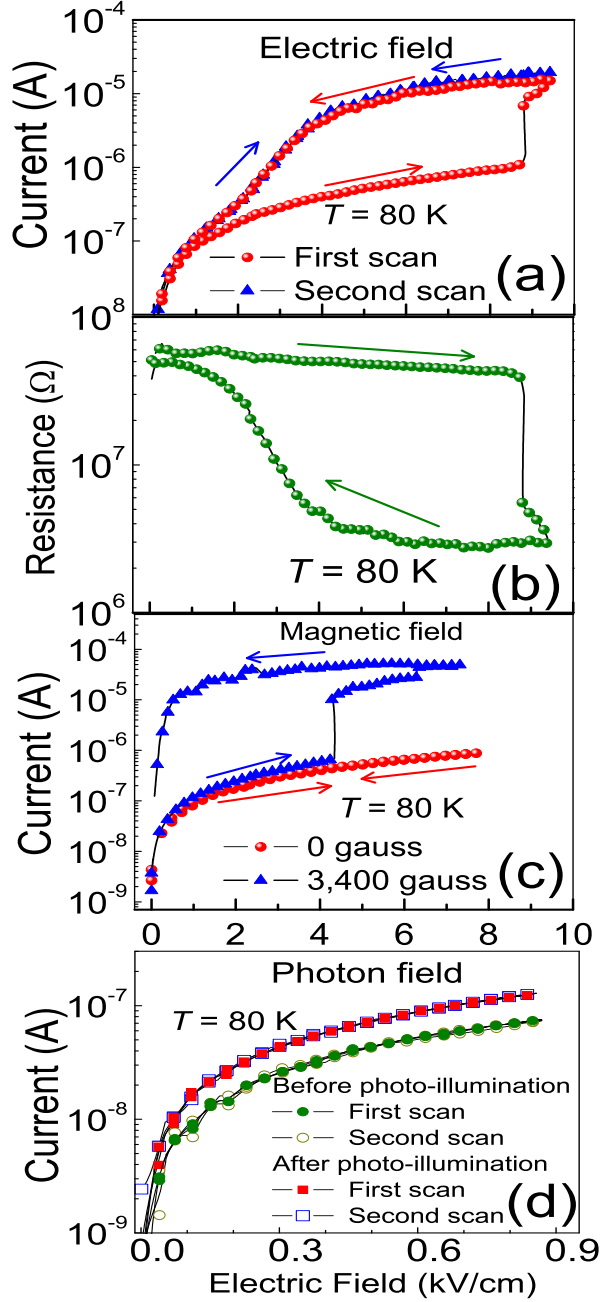}
 \caption{(Colour online) (a)
Two successive Current (\textit{i}) - Electric field(\textit{E})
characteristics of the sample in dark at 80 K. For the first scan
(filled circles), while increasing \textit{E}, a current switching
is observed at \textit{E$_{C}$} $\approx$ 9 kV/cm. The branch
corresponding to decreasing \textit{E} exhibit no reverse
switching. A second \textit{i-E} scan (filled triangles) traces a
curve similar to the curve corresponding to the decreasing
\textit{E} branch of the first scan indicating that sample
collapsed to a persistent low resistance state. (b) The electric
field dependence of the sample resistance for the first scan of
(a). (c) \textit{i-E} characteristics of the sample in dark at 80
K for 0 $\leq$ \textit{E} $<$ \textit{E$_{C}$} (filled circles)
showing no current switching. A second \textit{i-E} scan (filled
triangles) in dark with \textit{H} = 3,400 gauss applied parallel
to the plane of the film brings the switching field to $\approx$
4kV/cm. (d) A pair of \textit{i-E} scans of the sample in dark at
80 K for 0 $\leq$ \textit{E} $\ll$ \textit{E$_{C}$} (filled and
open circles respectively) followed by a pair of \textit{i-E} scan
after irradiating with 30 pulses of photon energy
(\textit{h$\nu$}) 2.33 eV and intensity 1 mJ/pulse (filled and
open squares respectively). } \label{fig2}
\end{figure}

Now we establish the complementarity of the photon flux and
electric field in inducing the I-M transition. In
Fig.~\ref{fig2}(a) the current (\textit{i}) - electric field
(\textit{E}) characteristics of the sample in dark at 80 K are
shown. For the first scan with increasing bias field the current
initially increases as \textit{i} $\sim$ \textit{E$^{ n}$},
\textit{n}$\approx$ 1.16. This small deviation from the ohmic
behaviour (\textit{n} = 1) is also seen in the \textit{R-E} plot
of Fig.~\ref{fig2}(b). However, when the electric field across the
sample exceeded a critical value \textit{E$_{C}$} ($\approx$ 9
kV/cm), the resistance drops abruptly by a factor of $\approx$ 10
indicating an electric field induced switching.  We denote the
maximum electric field applied across the sample as
\textit{E$_{max}$} and the corresponding resistance
\textit{R$_{min}$}. On decreasing \textit{E} the resistance
remains around \textit{R$_{min}$} for high electric fields and
then starts increasing when \textit{E} falls below $\approx$ 5
kV/cm. This \textit{E}-field induced metallic state is persistent
as is evident from the fact that a second \textit{i-E} scan
follows the decreasing \textit{E} branch of the first scan.
However before we attribute this spectacular effect to any
fundamental process such as field induced melting of the charge
ordered state, the role of simple Joule heating in triggering this
transition needs to be looked into. The following observations
suggest that the effects seen here is not a consequence of
heating; (1) Power dissipated in the circuit even at the peak
current is $<$ 1 mW. (2) The \textit{i-E} curves in repeated
cycles performed at the same temperature over a period of several
hours are ohmic with several orders of magnitude lower resistance.
This is not expected in the case of Joule heating in which case
the sample should come back to its virgin state when the source of
heat is removed. (3) Similar \textit{i-E} measurements on some
insulating non-charge ordered manganites LaMnO$_{3}$,
Sm$_{0.55}$Sr$_{0.45}$MnO$_{3}$ do not show current switching
effects. These observations clearly indicate that the electric
field and perhaps the associated current, but  not the heat
induces a structural/electronic phase change in the material,
which is healed only after heating the sample above its CO
temperature.

After heating the sample to 300 K and cooling it back to 80 K we did
another \textit{i-E} measurement in dark, as shown in
Fig.~\ref{fig2}(c) ( filled circles), but this time the electric
field was kept below \textit{E$_{C}$} ($\approx$ 9 kV/cm). As
expected, we did not see any \textit{E}-field induced switching. This
was followed by another \textit{i-E} measurement in the presence of a
magnetic field (\textit{H}) of 3,400 gauss applied in the plane of
the film. As seen in Fig.~\ref{fig2}(c), a sharp jump in the current
is observed at a field strength of $\approx$ 4 kV/cm, which is lower
than the zero-field value of the switching field (\textit{E$_{C}$}).
Yet again the high conductivity state is persistent. While the
observed I-M transition is not purely a magnetic field induced effect
as an assisting electric field was required, the application of the
magnetic field brings down the \textit{E$_{C}$} substantially.
 Finally, the sample was once again cooled to 80 K after heating it to
room temperature. Two successive \textit{i-E} scans  were made in
dark and \textit{E} $<$ \textit{E$_{C}$}  to ensure the stability of
the insulating state . After the end of the second scan, the sample
was irradiated with 30 pulses of 2.33 eV radiation  and the
\textit{i-E} scans were taken again, the result of which are shown in
Fig.~\ref{fig2}(d). As seen from the graph, there is a clear drop in
the resistance after photo-irradiation although this drop is much
smaller as compared to the drop at 30 K.

\begin{figure}
\includegraphics [width=12cm]{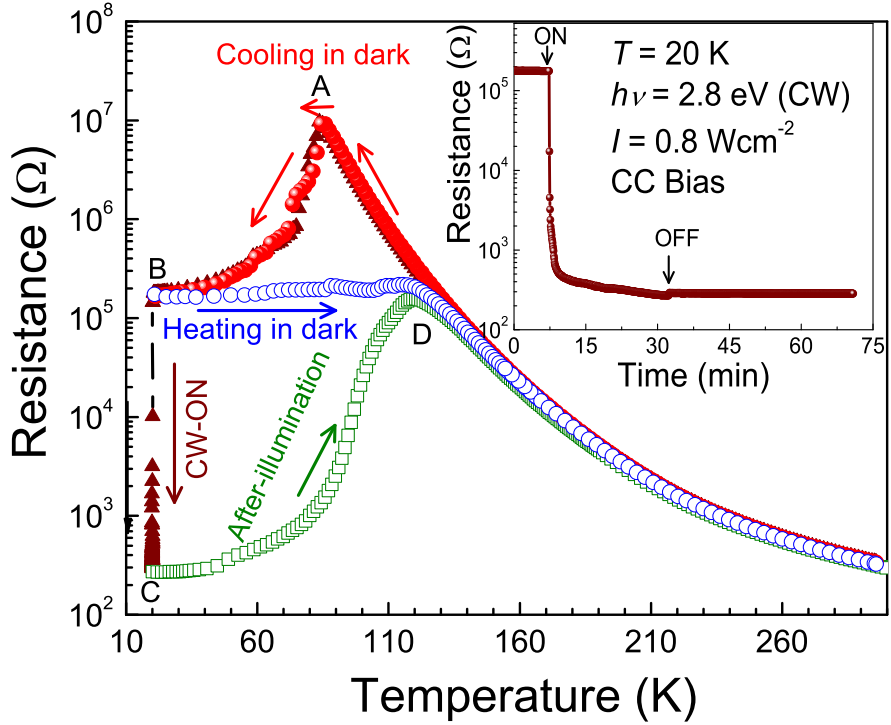}
\caption{(Colour online)
Temperature dependence of resistance measured in the Constant
Current (CC) mode while cooling (filled circles). The point A on
this curve corresponds to the onset of the \textit{E}-field
induced formation of metallic clusters that results in the
decrease of resistance. The resistance decreases down to the
lowest temperature (20 K) of the measurement indicated by the
point B and reaches a value of $\approx$ 0.15 M$\Omega$ at 20 K.
On heating the sample back to 300 K in dark (open circles) the
resistance approximately remains constant around 0.15 M$\Omega$
for \textit{T} $\leq$ 120 K. Above 120 K  the resistance reveals a
thermally activated behaviour. The sample was once again cooled to
20 K (filled triangles). This time at B a CW (\textit{h$\nu$} =
2.8 eV, \textit{I} = 0.8 Wcm$^{-2}$) laser beam is made incident
on the sample for $\approx$ 25 mins during which the resistance
drops and reaches point C as indicated on the curve. Finally, the
sample is heated back to 300 K in dark. The heating trace (open
squares) for \textit{T} $<$ 120 K reveal a metallic behaviour and
for \textit{T} $\geq$ 120 K the same thermally activated behaviour
was observed. The inset shows the time dependence of the sample
resistance with the laser on for about 25 mins and then switched
off. } \label{fig3}
\end{figure}

In Fig.~\ref{fig3} we show results of \textit{R-T} measurement in the
CC mode. For \textit{T} $\geq$ 85 K, the cooling curve (filled
circles) in dark is identical to the cooling curve of the CV mode
(Fig.~\ref{fig1}). Below 85 K, however, the behaviour of \textit{R-T}
is different from the CV data. Here the resistance starts decreasing
below \textit{T} $\approx$ 85 K and reaches a value of $\approx$ 0.15
M$\Omega$ at 20 K. This decrease can be attributed to a
voltage-induced melting of the CO state. Unlike the CV measurement,
the electric field across the sample in a CC measurement is not
constant.  It increases to ensure a constant current as the sample
resistance rises on lowering the temperature. At \textit{T} $\approx$
85 K, the field becomes comparable to \textit{E$_{C}$} and the
resistance exhibits a downturn. However, as \textit{R} goes down,
the sample bias voltage also decreases in order to maintain the CC
condition. While this feedback mechanism arrests a precipitous drop
in resistance, a gradual lowering of \textit{R} with decreasing
temperature continues since the threshold electric field
\textit{E$_{C}$} also decrease as the temperature goes down. Upon
heating the sample from 20 K, the resistance remains constant around
0.15 M$\Omega$ till 120 K and then falls following the cooling curve.
In a second \textit{R-T} measurement, the sample was cooled to 20 K
in dark (filled triangles) followed by photo-illumination with 2.8 eV
CW radiation of intensity 0.8 Wcm$^{-2}$ for $\approx$ 25 mins. The
laser radiation was then cut off and the resistance monitored as a
function of time for the next 40 mins while holding the temperature
at 20 K. As seen clearly, the photo-illumination decreases the
resistance by $\approx$ 3 orders of magnitude. It is evident that in
the CC measurements the switching transition occurs in two steps,
each leading to the conversion of a certain fraction of the material
to the metallic phase. The first state is initiated by the electric
field at $\approx$ 85 K . Although the stimulus acts uniformly across
the bridge, the breakdown presumably takes place along the weakest
path, thereby creating only a few percolating channels through which
majority of the current flows leaving most of the sample in the
original charge ordered insulating state. The second step is
photo-excitation of carriers across the charge-order  gap and
creation of new metallic patches which eventually coalesce and create
more percolative channels. In the inset of Fig.~\ref{fig3} we show
the time dependence of sample resistance at 20 K during exposure and
after exposure to 0.8 Wcm$^{-2}$ CW radiation of energy 2.8 eV.
During exposure the sample resistance drops, first abruptly and then
very slowly with time. Once the laser is switched off  no noticeable
change in the resistance is seen over an extended period of time
($\approx$ 2,400 s). This observation rules out the possibility that
the PIMT is caused by laser heating as $\sim$ 2,400 seconds is a long
enough time for the sample to thermalize. The similarity of the
curves obtained on heating the sample in dark from 20 K to 300 K
after photo-illumination at 20 K, and heating from 20 K to 300 K with
the laser beam on during the heating (not shown), further suggested
that thermal effects of laser irradiation are insignificant.

\begin{figure}
\includegraphics [width=12cm]{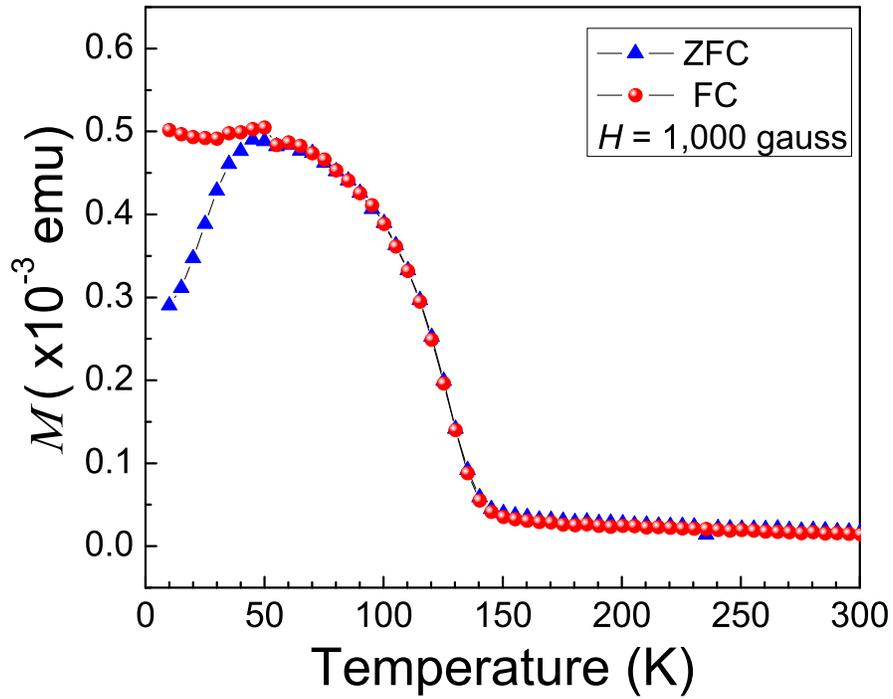}
\caption{(Colour online)
Temperature dependence of magnetization (\textit{M}) measured in a
in-plane field of 1,000 gauss for zero-field-cooling (ZFC) (filled
triangles) and field-cooling (FC) (filled circles). The rapid
increase of magnetization below 140 K marks the paramagnetic to
ferromagnetic transition. Below 50 K the FC branch exhibits a
saturation behaviour. }\label{fig4}
\end{figure}

The distinct peak in resistance at $\approx$ 120 K on warming the
sample after light exposure can be related to the magnetic
transition in the system.  Figure ~\ref{fig4} shows the
zero-field-cooled (ZFC) and field-cooled (FC) magnetization of the
sample measured between 10 K and 300 K. Although an absolute
measurement of the ordered moment per Mn site of the sample is
difficult in the present case due to the large paramagnetic
contribution from the substrate (NGO), the data clearly show onset
of spontaneous magnetization at \textit{T} $\approx$ 140 K. The
fraction of the ferromagnetic phase which is also metallic is,
however, not large enough to yield signature of this magnetic
transition in the \textit{R} vs \textit{T} plot. But on
photo-illumination, the phase become large enough to open a
percolating path. The peak in the resistivity at \textit{T}
$\approx$ 120 K can be attributed to  the ferromagnetic (FM) to
paramagnetic (PM) transition as seen in other manganites such as
La$_{0.7}$Ca$_{0.3}$MnO$_{3}$ \cite{coey}. The density of such
paths depends upon the photon flux as evident from the extent of
change in resistance on CW and pulsed laser irradiation.

\begin{figure}
\includegraphics [width=12cm]{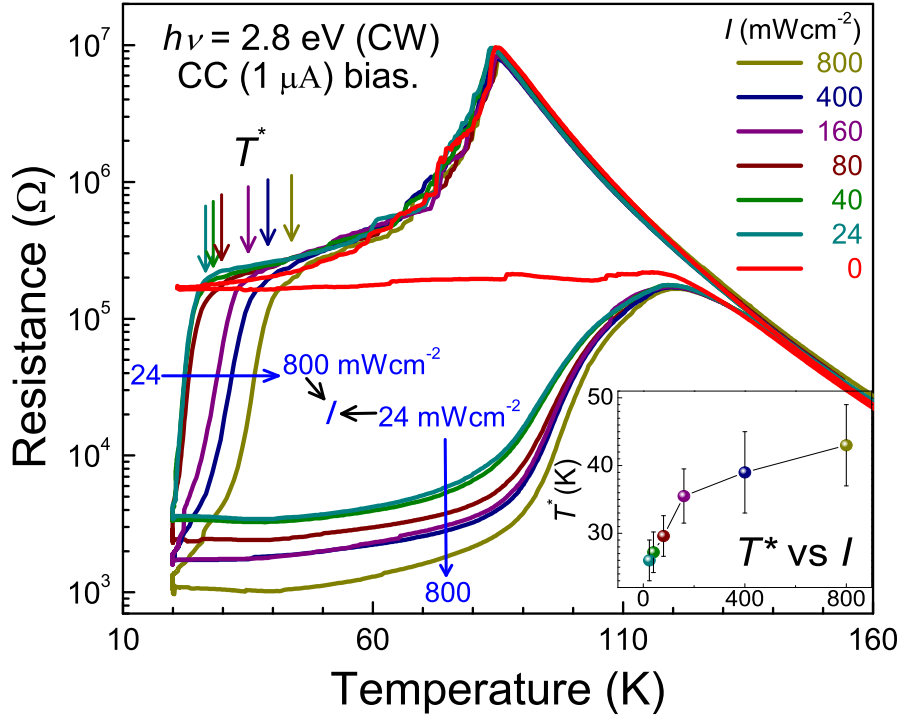}
\caption{(Colour online)
Temperature dependence of sample resistance is shown for different
intensities of 2.8 eV radiation measured in a Constant Current
(CC) bias mode. For each measurement, the sample was exposed to
laser illumination at 150 K while cooling from 300 K and remained
exposed till 20 K. At 20 K the sample was soaked in light for
$\approx$ 20 mins and finally heated back to 300 K in dark. The
temperature at which the resistance starts decreasing rapidly
(\textit{T$^{*}$}) (indicated by downward pointing vertical arrows
in the figure) increases with increasing intensity of incident
radiation (\textit{I}) as shown in the inset of the figure. Also,
the value of the resistance at 20 K depends on \textit{I} and is
lower for higher intensities consistent with the fact that at
higher intensity the number of photon penetrating deeper into the
sample increases.}\label{fig5}
\end{figure}

A further confirmation of this result comes from the dependence of
sample resistance on intensity (\textit{I}) of 2.8 eV CW radiation
measured in the CC mode as shown in Fig.~\ref{fig5}. For all the
scans shown in this figure, the exposure to light was initiated at
150 K while cooling from 300 K and continued till the sample
temperature reaches 20 K. During the heating cycle from 20 K to 300 K
the sample remained in dark. The \textit{R} vs \textit{T} curves
taken at different intensity of light remain the same down to
$\approx$ 50 K as the one taken in dark. However, a precipitous drop
in  resistance occurs at a critical temperature \textit{T$^{*}$} ($<$
50 K) which scales non-linearly with the photon flux, as shown in the
inset of Fig.~\ref{fig5}. In order to understand the data we must
recall that in a CC-mode measurement the changing electric field also
creates a metallic phase in addition to the conducting phase fraction
\textit{p} resulting from photo-irradiation. Since for the higher
photon flux, the induced metallic phase  is already large,  a small
contribution to \textit{p} from the electric field is required such
that the total $p$ reaches $p_{c}$, the percolation threshold. This
is the reason why \textit{T$^{*}$} moves up progressively with the
photon flux (inset Fig.~\ref{fig5}).  It is also interesting to note
that the resistance reached at 20 K on photo-exposure is
progressively smaller with the increasing photon flux.

It is interesting to note that  a somewhat similar IMT initiated
not by a CW  laser but a broad band xenon lamp has been seen in
oxygen deficient La$_{2/3}$Sr$_{1/3}$MnO$_{3}$  and
Pr$_{2/3}$Sr$_{1/3}$MnO$_{3}$ films\cite{gilabert,cauro2}. While
these authors have attributed  the IMT to photo induced
electron-hole pair generation, where electrons are trapped at
oxygen vacancies and the free hole contribute to conduction,
doubts about probable heating effects have been raised
\cite{cauro2}. While our experiments with pulsed laser light
clearly show that the photo-induced metallic state is persistent
(survives after removal of the light as long as the temperature is
not raised beyond \textit{T$_{CO}$}), we put this issue of sample
heating to rest by performing CW photoexposure experiments at the
highest intensity on fully oxygenated manganite thin films of
insulating ground state such as LaMnO$_{3}$ and
Sm$_{0.55}$Sr$_{0.45}$MnO$_{3}$. These films are of similar
geometry and thickness deposited on substrates (LaAlO$_{3}$ and
SrLaAlO$_{4}$) of similar thermal conductivity. We do not see any
perceptible drop in the resistance of the samples on
photoexposure. The temperature rise due to heating is only about
1.7 K/Watt cm$^{-2}$ at \textit{T} $\approx$ 76 K. The resistance
of the final state reached after photo-illumination with different
intensities and different durations of exposure can be understood
qualitatively by considering excitation of electrons from
Mn$^{3+}$ to Mn$^{4+}$ and thus a local disruption of the CO
state. We believe these disruptions first localize in the form of
small polarons, then grow into large polarons and finally form a
conducting channel with the increasing photon flux.

\begin{figure}
\includegraphics [width=12cm]{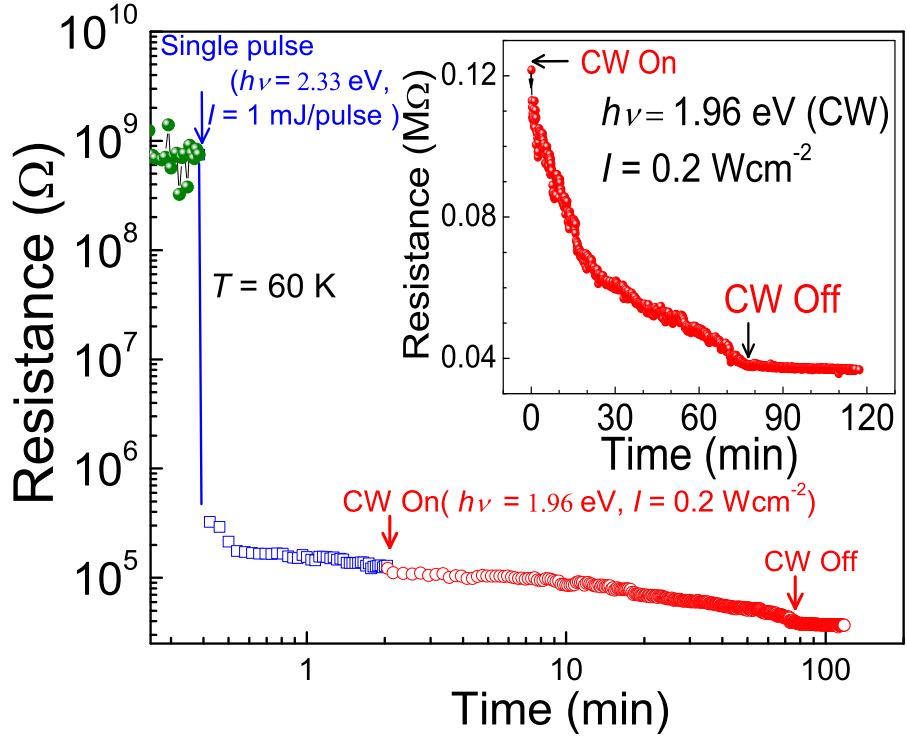}
\caption{(Colour online)
Resistance (\textit{R}) versus Time in minutes (\textit{t}) plot
of the sample at 60 K, in dark (filled circles) followed by
irradiation with a single 2.33 eV pulse of energy 1 mJ at
\textit{t} $\approx$ 0.35 min, which triggered an abrupt drop in
\textit{R}. The resistance was monitored continuously till
\textit{t} = 2 mins (open squares) during which it decreased at a
very slow rate. At \textit{t} = 2 mins, the sample was subjected
to CW illumination (\textit{h$\nu$} = 1.96 eV, Intensity $\approx$
0.2 Wcm$^{-2}$) and \textit{R} was continuously monitored for 2
$\leq$ \textit{t} $\leq$ 75 (open circles). No discernible change
in the resistance of the sample was seen on CW illumination as
seen from the inset where R for 2 $\leq$ \textit{t} $\leq$ 75 is
plotted on a linear scale.}\label{fig6}
\end{figure}

In Fig.~\ref{fig6} we show the effect of photo-illumination  with a
single shot of 2.33 eV ($\epsilon$ = 1 mJ/pulse), followed by
exposure to 1.96 eV CW (\textit{I} = 0.2 Wcm$^{-2}$) laser radiation
for about 75 mins. This measurement was performed in the CV  mode at
60 K. We note that the 2.33 eV pulse leads to a $\approx$ 4 orders of
magnitude drop in resistance and the subsequent CW-beam causes a
negligible further drop. Contrary to a previous  report
\cite{takubo}, no annealing effects of CW laser irradiation which
lead to the recovery of the original insulating state is seen at
these intensities.

\section{Conclusion}
In conclusion, we have observed a persistent photo-induced I-M
transition in \laprs thin films. Our observations point to the
fact that electric, magnetic and photon fields acting
independently induces metallic domains in the sample. A
complementary role of these stimuli is seen from the enhancement
of the effect when two of them are applied simultaneously.
Although these different perturbations couple to different
internal degrees of freedom, like the magnetic field couples to
the spin system where as light and electric fields couple to the
charge, a strong correlation among these  degrees of freedom
causes a collapse to the same final state. Moreover, this
particular manganite appears to be the only one discovered till
date where all the degrees of freedom  are susceptible to
perturbations of comparatively smaller magnitude. The localized
nature of the persistent photo-induced I-M transition in
conjunction with the fact that the metallic and the insulating
regions will have different refractive index and also different
Kerr rotation \cite{weisheit}, if they can be shown unambiguously
to be ferromagnetic due to the state of the metallic clusters,
make this material promising for magnetic and optical memory
application. The need of pump-probe experiments to see Kerr
rotation after photoexposure is important.

\acknowledgments This research has been supported by grants from
the Indo-French Center for the Promotion of Advanced Research New
Delhi and the Board of Research in Nuclear Sciences (BRNS)
Government of India. We thank R. Sharma for technical assistance
in magnetization measurements and Navneet Pandey for depositing
films of LaMnO$_{3}$ and Sm$_{0.55}$Sr$_{0.45}$MnO$_{3}$ which
were used to estimate the heating effect. 

\clearpage
\clearpage
\clearpage
\clearpage
\clearpage
\clearpage

\end{document}